\documentclass[preprint,showpacs,preprintnumbers,amsmath,amssymb]{revtex4}
\usepackage{graphicx}
\usepackage{dcolumn}
\usepackage{bm}
\usepackage{color}

\font\scripti=cmmi7
\font\scriptscripti=cmmi5
\def\sib#1{\setbox0 = \hbox{\scripti #1}
  \kern-.02em\copy0\kern-\wd0
  \kern.04em\box0} 
\def\ssib#1{\setbox0 = \hbox{\scriptscripti #1}
  \kern-.02em\copy0\kern-\wd0
  \kern.04em\box0} 
\font\tenib=cmmib10 
\skewchar\tenib='177 \skewchar\tenib='177 \skewchar\tenib='177
\def\pbold#1{\setbox0 = \hbox{$ #1 $}
  \kern-.022em\copy0\kern-\wd0
  \kern.011em\copy0\kern-\wd0
  \kern.011em\copy0\kern-\wd0
  \kern.011em\copy0\kern-\wd0
  \kern.011em\box0} 

\usepackage{graphicx}
\usepackage{dcolumn}
\usepackage{bm}

\def\lesssim{\ \raise.3ex\hbox{$<$}\kern-0.8em\lower.7ex\hbox{$\sim$}\ }
\def\gesim{\ \raise.3ex\hbox{$>$}\kern-0.8em\lower.7ex\hbox{$\sim$}\ }

\begin{document}
\title{QCD-like phase diagram with Efimov trimers and Cooper pairs in resonantly interacting SU(3) Fermi gases }
\author{Hiroyuki Tajima and Pascal Naidon}
\affiliation{Quantum Hadron Physics Laboratory, RIKEN Nishina Center (RNC), Wako, Saitama, 351-0198, Japan}
\date{\today}
\begin{abstract}
We investigate color superfluidity and trimer formation in resonantly interacting SU(3) Fermi gases with a finite interaction range.
The finite range is crucial to avoid the Thomas collapse and treat the Efimov effect occurring in this system.
Using the Skorniakov-Ter-Martirosian (STM) equation with medium effects, we show the effects of the atomic Fermi distribution on the Efimov trimer energy at finite temperature. 
We show the critical temperature of color superfluidity within the non-selfconsistent $T$-matrix approximation (TMA).
In this way, we can provide a first insight into the phase diagram as a function of the temperature $T$ and the chemical potential $\mu$.
This phase diagram consists of trimer, normal, and color-superfluid phases,
and is similar to that of quantum chromodynamics (QCD) at finite density and temperature, indicating a possibility to simulate the properties of such extremely dense matter in the laboratory.
\end{abstract}
\maketitle
\section{Introduction}
\label{sec1}
The concept of quantum simulation has opened new possibilities
for exploring the properties of novel materials as well as exotic matter in extreme conditions \cite{Bloch,Georgescu,Gross}.
Recently, ultracold atoms have been used as quantum simulators of strongly correlated systems thanks to the tunability of their physical parameters such as interparticle interaction \cite{Dalfovo,Bloch2,Giorgini,Chin}.
For example, ultracold atomic Fermi gases loaded into an optical lattice can realize the Hubbard model, which is relevant to high-$T_{\rm c}$ cuprate superconductors \cite{Gross,Bloch3,Kohl,Jordens,Mazurenko}.
The antiferromagnetic behavior which plays an important role for the superconducting mechanism of high-$T_{\rm c}$ cuprates
has been observed in this atomic system \cite{Gross,Mazurenko}.
As another example, strongly interacting two-component Fermi gases can be used as quantum simulators of dilute neutron matter \cite{Gezerlis,vanWyk,Strinati}.
Indeed, the observed thermodynamic quantities of homogeneous Fermi gases near the unitarity limit \cite{Navon,Horikoshi,Tajima2,HorikoshiE}
quantitatively reproduce the equation of state of neutron matter obtained by numerical simulations \cite{APR} which is crucial for understanding the interior of a neutron star \cite{Oertel,Baym}.
\par
The analog simulation of quantum chromodynamics (QCD) \cite{Greiner,Fukushima}, where quarks with three colors strongly interact with each other, can be regarded as one of the next important challenges for cold-atom physics \cite{Maeda,Cirac,Ohara}.
While numerous efforts have been made to explore the phase diagram of finite-density QCD where various phenomena such as color superconductivity \cite{Alford} have been theoretically proposed,
they could not be confirmed by high-energy experiments due to the extreme densities, nor by exact numerical simulations due to the sign problem. 
On the other hand, three-component fermion systems have been experimentally realized 
by using mixtures of fermionic atoms in three different internal states
\cite{Ohara,Ottenstein,Huckans,Williams,Spiegelhalder,Lompe,Wenz,Nakajima1,Nakajima2}.
These systems reach the quantum degenerate regime around $T\simeq 0.3T_{\rm F}$ \cite{Ottenstein} where $T_{\rm F}$ is the Fermi degenerate temperature of non-interacting atoms.
In this regime, the existence of color superfluidity similar to color superconductivity in QCD has been theoretically examined in three-component Fermi gases \cite{He,Rapp,Paananen,Bedaque,Kantian,Catelani,Martikainen,Ozawa}.
Although there are several differences between these atomic systems and QCD,
they constitute a good starting point to study the strong-coupling effects in three-component fermion systems.
\par
In most of the previous works on color superfluidity \cite{He,Rapp,Paananen,Bedaque,Kantian,Catelani,Martikainen,Ozawa},
zero-range interactions have been used to describe the attraction between the three components. 
However, such zero-range interactions are known to lead to a collapse of the system   
\cite{Thomas,Blume} in connection with the existence of Efimov trimers \cite{Efimov1,Efimov2,Braaten,Naidon2,Greene,Williams,Nakajima1,Nakajima2}.
A finite range of interactions is therefore necessary to prevent the collapse
and properly treat Efimov trimers.
The effect of finite range and the Efimov trimers on the phase diagram of three-component fermionic system was studied in Ref.~\cite{Nishida}.
However, this study was limited to zero temperature.
\par
In this paper, we investigate in-medium Efimov trimers and color superfluidity in resonantly interacting SU(3) Fermi gases with finite interaction range and finite temperature. By including medium effects in the Skorniakov-Ter-Martirosian (STM) equation \cite{STM} which exactly describes Efimov physics in a three-body problem \cite{Naidon3}, we discuss how the trimer energy is changed at finite temperature.
To explore many-body physics related to color superfluidity, we employ the non-selfconsistent $T$-matrix approximation (TMA) \cite{Pieri,Perali1,Perali2,Tsuchiya,Tsuchiya2,Watanabe,Perali3,Mueller1,Palestini,Ota,Tajima3}, which can correctly describe the Bardeen-Cooper-Schrieffer to Bose-Einstein-condensation (BCS-BEC) crossover \cite{Eagles,Leggett,NSR,SadeMelo,Ohashi,Chen} in a strongly interacting two-component $^{40}$K \cite{Regal} and $^6$Li \cite{Zwierlein} Fermi gas.
Combining the results of these two approaches, we obtain the phase diagram with respect to the temperature $T$ and the chemical potential $\mu$.
\par
This paper is organized as follows.
In Sec. \ref{sec2}, we explain the Hamiltonian for resonant SU(3) Fermi gases and our framework of in-medium STM equation and non-selfconsistent TMA.
In Sec. \ref{sec3}, we show our numerical results of the in-medium Efimov trimer energy and discuss the finite temperature behavior of them, as well as the color-superfluid instability.
Finally, we summarize this paper in Sec. \ref{sec4}.
For simplicity, we take $\hbar=k_{\rm B}=1$ and the system volume $V$ is taken to be unity.

\section{Formalism}
\label{sec2}
\subsection{Hamiltonian}
We consider a symmetric three-component fermionic system.
We model the interaction between two different components by a two-channel Feshbach resonance model \cite{Chin,Nishida} consisting of an open channel and a closed channel described by a single diatomic molecular state.
We neglect the interaction within the open channel which corresponds 
to the limit of closed-channel dominated resonances.
Furthermore, the coupling between open channel and molecular states is taken as a contact-type coupling.
The three-component fermionic system is therefore described by the following Hamiltonian,
\begin{eqnarray}
\label{eq1}
H&=&\sum_{j=1,2,3}\sum_{\bm{p}}\xi_{\bm{p}}^{\rm F}c_{\bm{p},j}^{\dag}c_{\bm{p},j}+\sum_{i<j}\sum_{\bm{q}}\xi_{\bm{q}}^{\rm B}b_{\bm{q},ij}^{\dag}b_{\bm{q},ij} \cr
&&+g\sum_{i<j}\sum_{\bm{p},\bm{q}}\left(b_{\bm{q},ij}^{\dag}c_{-\bm{p}+\bm{q}/2,i}c_{\bm{p}+\bm{q}/2,j}+{\rm h. c. }\right),
\end{eqnarray}
where $\xi_{\bm{p}}^{\rm F}=\frac{\bm{p}^2}{2m}-\mu$ and $\xi_{\bm{q}}^{\rm B}=\frac{\bm{q}^2}{4m}+\nu-2\mu$ are the kinetic energies of a fermion with mass $m$ and the diatomic molecule, respectively, and $\mu$ is the fermionic chemical potential.
$c_{\bm{p},j}$ and $b_{\bm{q},ij}$ are annihilation operators of a Fermi atom with the internal state $j=1,2,3$ and a diatomic molecule made of $i-j$ atomic pair.
The energy of diatomic molecules $\nu$ and the atom-dimer Feshbach coupling $g$ are related to the scattering length $a$ and the range parameter $R_{*}$ as follows,
\begin{eqnarray}
\label{eq2}
\frac{m}{4\pi a}= -\frac{\nu}{g^2}+\sum_{\bm{p}}\frac{m}{\bm{p}^2},
\end{eqnarray}
\begin{eqnarray}
\label{eq3}
R_{*}=\frac{4\pi}{m^2g^2}.
\end{eqnarray}
We note that the effective range $r_{\rm e}$ of this two-channel interaction is here negative and is associated with $R_*=-\frac{1}{2}r_{\rm e}$.
In this paper, we focus on the unitarity limit $1/a=0$.
\par

\subsection{non-selfconsistent $T$-matrix approximation}
\begin{figure}[t]
\begin{center}
\includegraphics[width=8.5cm]{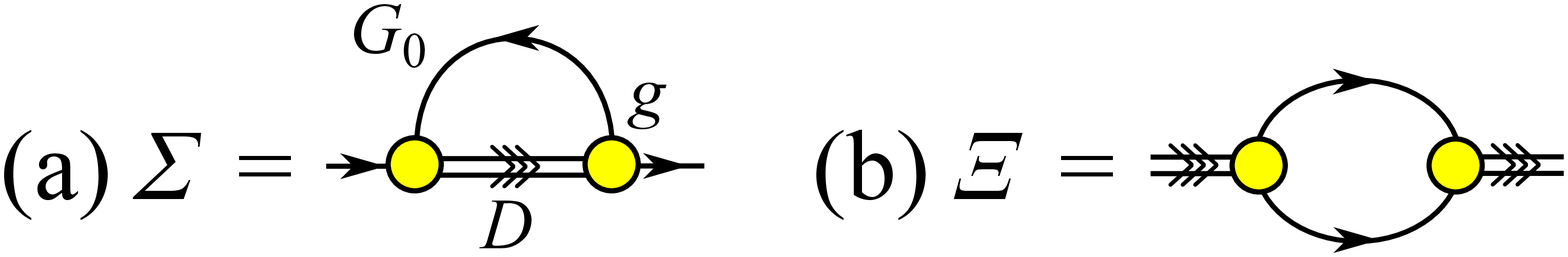}
\end{center}
\caption{Feynman diagrams of the self-energies for (a) a Fermi atom $(\Sigma)$ and (b) a diatomic molecule $(\Xi)$.
Single and double solid lines show the thermal Green's function of the bare Fermi atom and the dressed diatomic molecule, respectively. Circles represent the Feshbach coupling $g$.
}
\label{fig2}
\end{figure}
We first employ the non-selfconsistent $T$-matrix approximation within the Matsubara formalism \cite{Pieri,Perali1,Perali2,Tsuchiya,Tsuchiya2,Watanabe,Perali3,Mueller1,Palestini,Ota,Tajima3} to determine thermodynamic properties such as critical temperature $T_{\rm c}$ of color superfluidity.
In this framework, the thermal Green's function of dressed atoms is given by
\begin{eqnarray}
\label{eq7}
G(p)=\frac{1}{i\omega_n-\xi_{\bm{p}}^{\rm F}-\Sigma(p)},
\end{eqnarray}
where we used the four-momentum notation $p=(\bm{p},i\omega_n)$ and $\omega_n=(2n+1)\pi T$ is the fermionic Matsubara frequency.
$\Sigma(p)$ is the fermionic self-energy. 
In the case of just two fermions in vacuum, the self-energy is given by the diagram shown in Fig. \ref{fig2}(a).
Retaining only this diagram, we obtain \cite{Tajima3}
\begin{eqnarray}
\label{eq8}
\Sigma(p)=2g^2T\sum_{Q}D(Q)G_0(Q-p),
\end{eqnarray}
where $Q=(\bm{Q},i\nu_{n'})$ and $\nu_{n'}=2\pi n'T$ is the bosonic Matsubara frequency.
$G_0(p)=1/\left(i\omega_n-\xi_{\bm{p}}^{\rm F}\right)$ and $D(Q)$ are the in-medium Green's functions of a non-interacting fermion and a dressed molecule, respectively.
We note that although this equation (\ref{eq8}) as the same form as that in vacuum,
here $G_0$ and $D$ contain the medium effects.
We also note that the factor $2$ in Eq. (\ref{eq8}) comes from the degree of freedom with respect to internal states.
\par
Similarly, the thermal Green's function of dressed molecules with the ultraviolet renormalization is given by
\begin{eqnarray}
\label{eq9}
D(Q)=\frac{1}{i\nu_{n'}-\xi_{\bm{Q}}^{\rm B}-\Xi(Q)},
\end{eqnarray}
where $\Xi(Q)$ is the bosonic self-energy diagrammatically shown in Fig. \ref{fig2}(b).
Here again, we take the vacuum form  
\begin{eqnarray}
\label{eq10}
\Xi(Q)&=&-g^2T\sum_{p}G_0(p+Q)G_0(-p) \cr
&=&g^2\sum_{\bm{p}}\frac{1-f(\xi_{\bm{p}+\bm{Q}}^{\rm F})-f(\xi_{\bm{p}}^{\rm F})}{i\nu_{n'}-\xi_{\bm{p}+\bm{Q}}^{\rm F}-\xi_{\bm{p}}^{\rm F}},
\end{eqnarray}
where $f(\xi)=1/(e^{\xi/T}+1)$ is the Fermi-Dirac distribution function.
The chemical potential $\mu/\varepsilon_{\rm F}$ with fixed number density $n$ [where $\varepsilon_{\rm F}=(2\pi^2 n)^{2/3}/(2m)$ is the Fermi energy of an ideal Fermi gas] is obtained by solving the number equation,
\begin{eqnarray}
\label{eq11}
n=3T\sum_{p}G(p)-6T\sum_{Q}D(Q).
\end{eqnarray}
In addition, we obtain the critical temperature $T_{\rm c}$ from the Thouless criterion $D^{-1}(\bm{Q}=\bm{0},i\nu_n=0)=0$ \cite{Thouless}, which gives ($1/a=0$)
\begin{eqnarray}
\label{eq12}
\frac{m^2\mu R_*}{2\pi}+\sum_{\bm{p}}\left[\frac{1}{2\xi_{\bm{p}}^{\rm F}}{\rm tanh}\left(\frac{\xi_{\bm{p}}^{\rm F}}{2T_{\rm c}}\right)-\frac{m}{\bm{p}^2}\right]=0.
\end{eqnarray}

\subsection{in-medium Skorniakov-Ter-Martirosian equation}
\begin{figure}[t]
\begin{center}
\includegraphics[width=8.5cm]{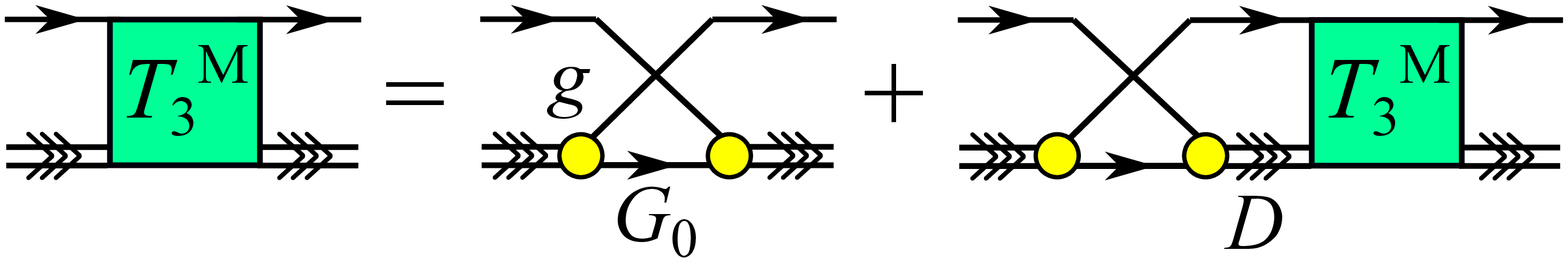}
\end{center}
\caption{Feynman diagram for the three-body $T$-matrix $T_3^{\rm M}$, where single and double solid lines represent the Green's function of a bare atom $G_0$ and a dressed molecule $D$, respectively.
Circles represent the Feshbach coupling $g$. 
}
\label{fig1}
\end{figure}
To determine the trimer energy $E_{3}^{\rm M}$ in the medium, we consider the three-body $T$-matrix equation \cite{Brodsky,Iskin} which is diagrammatically shown in
Fig. \ref{fig1}.
In the vacuum case, it gives the so-called Skorniakov-Ter-Martirosian (STM) equation which exactly describes Efimov physics in three-body problems.
Using the thermal Green's functions within the Matsubara formalism, we obtain the in-medium three-body $T$-matrix
\begin{eqnarray}
\label{eq4}
T_{3}^{\rm M}(p,p';P)&=&-g^2G_0(P-p-p')\cr
&&-2g^2T\sum_{q}G_0(P-p-q)G_0(q)D(P-q)T_3^{\rm M}(q,p';P).
\end{eqnarray}
We note that $p$, $p'$, and $P$ are the four-momenta of incoming fermion, outcoming fermion, and the center-of-the-mass, respectively (we suppress the index of internal states for simplicity).
The factor $2$ in the second term of r. h. s. of Eq. (\ref{eq4}) comes from the degree of freedom with respect to internal states.
$G_0(q)$ with $q=(\bm{q},i\Omega_n)$ indicates intermediate states of fermions which are integrated.
By including Pauli-blocking effects on $G_0(p)$ (see Appendix \ref{ApA}), we obtain 
the in-medium STM equation in the unitarity limit $(1/a\rightarrow 0)$,
\begin{eqnarray}
\label{eq5}
\left[-R_*\kappa(\bm{p})^2+4\pi\sum_{\bm{k}}\left\{\frac{F(\bm{k},\bm{p})}{\bm{k}^2+\kappa(\bm{p})^2}-\frac{1}{\bm{k}^2}\right\}\right]L\left(\bm{p}\right)=-8\pi\sum_{\bm{q}}\frac{F(\bm{q},\bm{p})L(\bm{q+\bm{p}/2})}{\bm{q}^2+\kappa(\bm{p})^2},
\end{eqnarray}
where $\kappa(\bm{q})^2=\frac{3}{4}\bm{q}^2-mE_3^{\rm M}$, $L(\bm{p})$ is the function defined by Eq. (\ref{eqA6}) in Appendix \ref{ApA}, and  
\begin{eqnarray}
\label{eq6}
F(\bm{p},\bm{q})=1-f(\xi_{\bm{p}+\bm{q}/2}^{\rm F})-f(\xi_{\bm{p}-\bm{q}/2}^{\rm F})
\end{eqnarray}
is the statistical factor associated with the Fermi-Dirac distribution of atoms given by $f(\xi_{\bm{p}}^{\rm F})=1/\left(e^{\xi_{\bm{p}}^{\rm F}/T}+1\right)$.
One can find that the ordinary STM equation in a three-body problem is recovered by setting $F(\bm{p},\bm{q})=1$.
We note that Eq. (\ref{eq6}) is qualitatively consistent with the previous works \cite{Niemann,Tajima1}.
We note however that Ref. \cite{Niemann} is restricted to $T=0$ with the choice
\begin{eqnarray}
\label{eqNiemann}
F(\bm{p},\bm{q})=\theta(\xi_{\bm{p}+\bm{q}/2}^{\rm F})\theta(\xi_{-\bm{p}+\bm{q}/2}^{\rm F}),
\end{eqnarray}
instead of Eq. (\ref{eq6}).
Here $\theta(x)$ denotes the Heaviside step function.
This choice allows the possibility of positive-energy solution of STM equation, called Cooper triples.
In contrast, our result Eq. (\ref{eq6}) (see Appendix \ref{ApA}) does not allow such solutions and we focus on the region where $E_{3}^{\rm M}\leq 0$. 
We briefly note that similar in-medium three-body equations were employed in nuclear physics~\cite{Beyer2,Beyer,Barbieri,Beyer3,Kuhrts,Pepin,Kvinikhidze,Mattiello}.
\par
\section{Results}
\label{sec3}
\begin{figure}[t]
\begin{center}
\includegraphics[width=8.5cm]{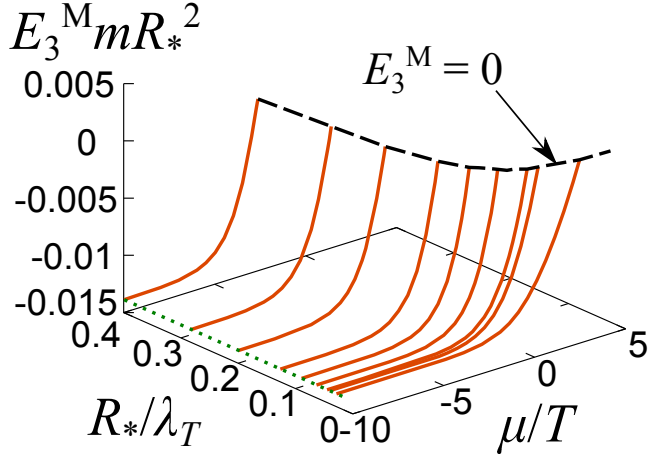}
\end{center}
\caption{The in-medium trimer energy $E_{3}^{\rm M}$ as a function of dimensionless quantities, $R_*/\lambda_T$ and $\mu/T$. 
$\lambda_T=\sqrt{\frac{2\pi}{mT}}$ is the thermal de Broglie wavelength.
The dotted line represents that in the vacuum limit given by $E_{3}^{\rm V}=-0.01385/(mR_*^2)$. 
}
\label{fig3}
\end{figure}
Figure \ref{fig3} shows the ground-state trimer energy $E_{3}^{\rm M}$ with the medium effects,
which can be obtained numerically from Eq. (\ref{eq5}) as a dimensionless function
\begin{eqnarray}
\label{eq13}
E_3^{\rm M}=\frac{1}{mR_*^2}X\left(R_*/\lambda_T,\mu/T\right),
\end{eqnarray}
where $\lambda_T=\sqrt{\frac{2\pi}{mT}}$ is the thermal de Broglie wavelength.
Physically, the range parameter $R_*$ gives the typical size of the Efimov trimer \cite{Petrov}.
In this regard, the ratio between $R_*$ and $\lambda_T$ represents how trimer states are affected by finite temperature effects.
In the case of $R_*\ll \lambda_T$ and $\mu/T\lesssim 0$, $E_{3}^{\rm M}$ approaches the vacuum limit given by $E_{3}^{\rm V}=-0.01385/(mR_*^2)$ which is close to that of a universal trimer \cite{Gogolin},
since the trimer size is small enough compared to the typical thermal length scale.
 In addition, the ratio $\mu/T$ is associated with the fugacity $z=e^{\mu/T}$ and represents Pauli-blocking effects due to the atomic Fermi distribution, which plays a significant role when $\mu>0$.
The absolute value of $E_{3}^{\rm M}$ is greatly reduced in the Fermi degenerate region.
Finally, $E_3^{\rm M}$ disappears in the region where $\mu/T$ and $R_*/\lambda_T$ are relatively large.
The physical interpretation of these effects is that  Fermi atoms from the medium weaken the Efimov attraction between three atoms forming a trimer state.
This phenomenon is somewhat similar to the Gor'kov-Melik-Barkhudarov (GMB) correction in weak-coupling superconductors for which the size of Cooper pairs is large \cite{GMB} and the pairing interaction is screened by the medium \cite{Floerchinger,Yu,Ruan,Pisani1,Pisani2}. 
\begin{figure}[t]
\begin{center}
\includegraphics[width=8.5cm]{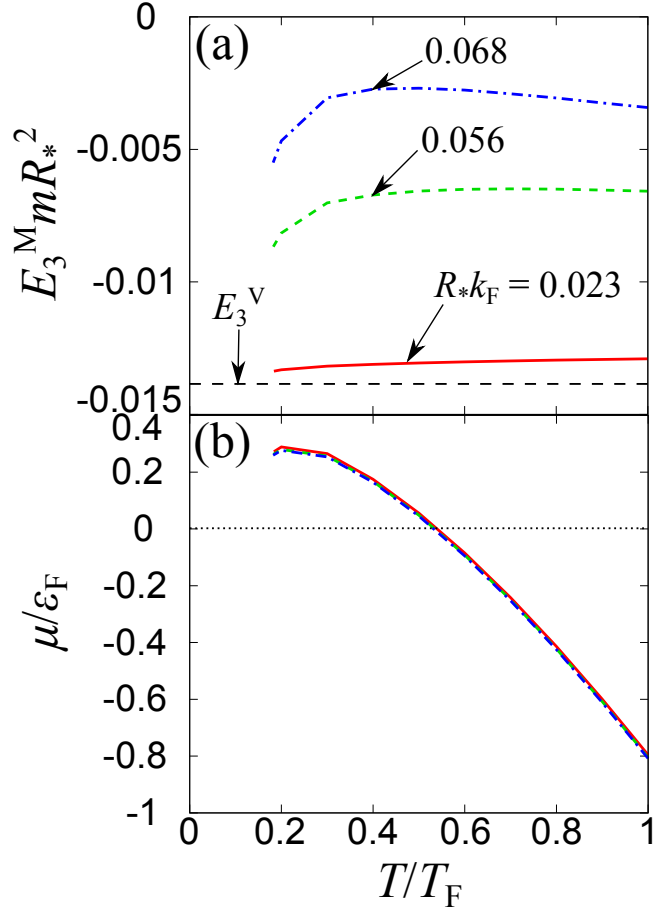}
\end{center}
\caption{(a) Trimer energy $E_3^{\rm M}$ with medium effects as a function of $T/T_{\rm F}$
obtained from the chemical potential $\mu/\varepsilon_{\rm F}$ calculated within the non-selfconsistent $T$-matrix approximation (TMA) above $T_{\rm c}$.
$T_{\rm F}$, $k_{\rm F}$, and $\varepsilon_{\rm F}$ are the Fermi degenerate temperature, the Fermi momenta, and the Fermi energy of ideal Fermi gases, respectively.
}
\label{fig4}
\end{figure}
\par
To see how these effects would appear in actual experiments at given temperatures and densities, we plot in
Fig. \ref{fig4} (a) the typical temperature dependence of trimer energy $E_3^{\rm M}$ at different range parameters
according to the density equation of state obtained from the non-selfconsistent $T$-matrix approximation (TMA).
By numerically solving the number equation Eq. (\ref{eq11}), we obtain the temperature-dependent chemical potential $\mu$ as shown in Fig. \ref{fig4} (b).
We then use $\mu$ as an input for in-medium STM equation given by Eq. (\ref{eq5}).
This gives $E_{3}^{\rm M}$ showing in Fig. \ref{fig4} (a).
$E_{3}^{\rm M}$ has a peak structure which can be understood from two effects,
that is, the evolution of the Fermi chemical potential and the decrease of $\lambda_T$ compared to the trimer size.
In the high-temperature limit, we can neglect the interaction effects and $\mu$ is given by the number equation of ideal gases $n=3\sum_{\bm{p}}e^{-\xi_{\bm{p}}^{\rm F}/T}$.
This gives approximately $\mu \simeq -\frac{3}{2}T{\rm ln}T$ and for large temperatures
$E_{3}^{\rm M}$ reproduces the vacuum result $E_{3}^{\rm V}$ due to the large negative $\mu$.
Decreasing the temperature makes the density and temperature effects more visible which weakens the trimer ($E_{\rm 3}^{\rm M}$ increases).
However, at very low temperature $\mu$ becomes almost constant while $\lambda_T$ increases as $1/\sqrt{T}$ with decreasing temperature. 
As a result, it becomes larger and larger compared to trimer size $R_*$ (which is fixed in this figure), which suppresses the temperature effects and the trimer strengthens ($E_{\rm 3}^{\rm M}$ decreases).  
This decrease becomes sharper with increasing $R_*$ as shown in Fig. \ref{fig4} (a).
We note that our calculation is stopped at $T=T_{\rm c}$,
where Eq. (\ref{eq11}) is invalid below $T_{\rm c}$ due to the existence of the superfluid gap.  
\begin{figure}[t]
\begin{center}
\includegraphics[width=8.5cm]{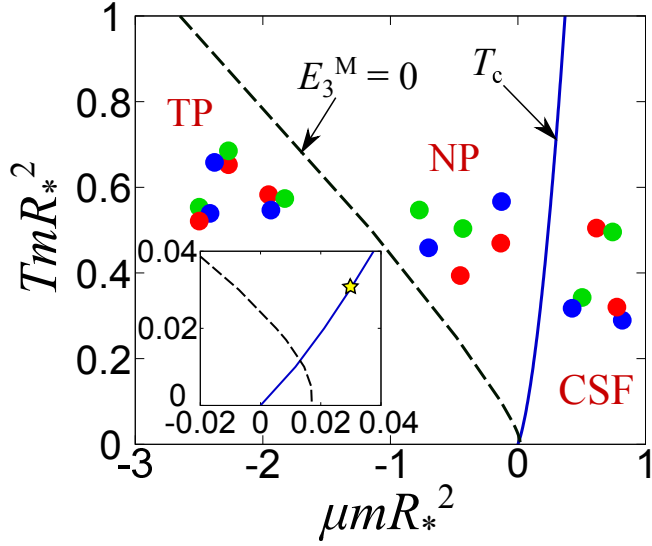}
\end{center}
\caption{Phase diagram of resonant SU(3) Fermi gases with the finite range parameter $R_*$.
The dashed and solid lines indicate the threshold of in-medium trimers $E_3^{\rm M}=0$ and the critical temperature of color superfluidity (CSF) $T_{\rm c}$, respectively.
$E_3^{\rm M}=0$ is regarded as the crossover boundary between trimer phase (TP) and normal phase (NP).
The inset shows a magnification around $\mu mR_*^2=0$.
The star indicates the parameters ($\mu mR_*^2=0.0298$ and $T mR_*^2=0.0302$) used in the calculation of the single-particle spectral function in Fig. \ref{fig6}. 
}
\label{fig5}
\end{figure}
\par
\begin{figure}[t]
\begin{center}
\includegraphics[width=8.5cm]{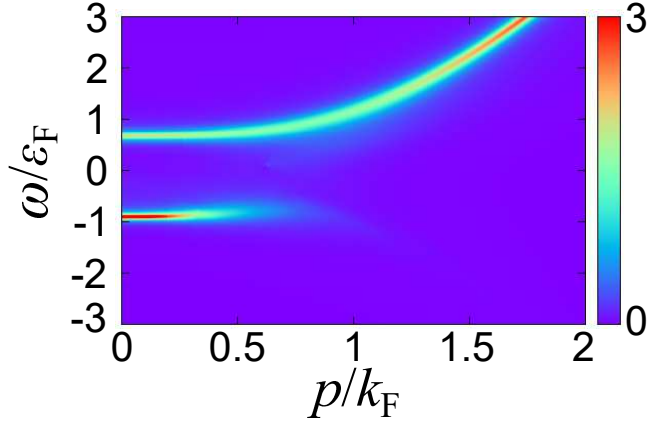}
\end{center}
\caption{Single-particle spectral function $A(\bm{p},\omega)$ in the arbitrary unit just above $T_{\rm c}$ with $\mu mR_*^2=0.0298$ and $T mR_*^2=0.0302$ indicated in the inset of Fig. \ref{fig5}.
}
\label{fig6}
\end{figure}

From these results, we get a first insight of the phase diagram with respect to the chemical potential $\mu$ and temperature $T$, as shown in Fig. \ref{fig5}.
We expect a trimer phase (TP) where all of the atoms are bound into ground-state Efimov trimers,
 a color superfluid phase (CSF) where all three kinds of pairs are condensed,
and a normal phase (NP) where the atoms form neither trimers nor condensed pairs. 
The boundary between TP and NP is estimated by the curve where $E_3^{\rm M}=0$.   
Threshold curve $E_3^{\rm M}=0$ does not mean any phase transition between trimer and normal phases
but it should be a good indication of how the trimer character disappears in the high-density region of this system. 
CSF is defined by the region below $T_{\rm c}$.
We note that $T_{\rm c}$ approaches $0$ in low-density (zero-range) limit at $\mu mR_*^2=0$ \cite{Tajima1}.
We also note that the high density limit of the critical temperature $T_{\rm c}^{\rm HDL}=0.133T_{\rm F}$ is simply obtained from the Bose-Einstein condensation (BEC) temperature of diatomic molecules in the presence of thermal-excited fermions and $\mu$ approaches $0$ in this limit \cite{Tajima1}.
This indicates that the system undergoes a crossover from unitary Cooper pairs to BEC of closed channel molecules.
This behavior is specific to the narrow-resonance two-channel model used in this work. 
Interestingly, Fig. \ref{fig6} is similar to the phenomenological phase diagram of QCD consisting of the hadron phase (analogue of TP), the color superconducting phase (analogue of CSF), and the deconfined quark phase (analogue of NP) \cite{Fukushima}.
However, while the phase transition at $T=T_{\rm c}$ between CSF and NP is of the second order, that of color superconductivity is of the first order due to the gauge coupling \cite{Alford}.
Moreover, the conjectured BEC-BCS crossover \cite{AbukiNishida} in QCD with increasing the chemical potential is opposite to the BCS-BEC crossover found in this model at $1/a=0$.
We stress again this is a particularity of the narrow-resonance two-channel model.
\par
Although our in-medium STM equation cannot be justified near $T=T_{\rm c}$ where the Bose-Einstein distribution of diatomic molecules increases at the low energy region,
one can find a quantum-phase-transition-like behavior around $\mu mR_*^2=0$.
The inset of Fig. \ref{fig5} shows the magnification around $\mu mR_*^2=0$, where the two curves we calculated cross each other.
In reality, the region near this point is expected to be dominated by strong multi-body correlation due to the competition between trimer formation and color superfluidity \cite{Pepin},
which cannot be captured by our treatment.
\par
Finally, we look at pairing fluctuations above $T_{\rm c}$.
Indeed, it is known that pairing fluctuations become strong near $T_{\rm c}$
in two-component Fermi gases near the unitarity limit.
Figure~\ref{fig6} shows the single-particle spectral function
\begin{eqnarray}
\label{eq14}
A(\bm{p},\omega)=-\frac{1}{\pi}{\rm Im}G(\bm{p},i\omega_n\rightarrow \omega+i\delta)
\end{eqnarray}
obtained from the analytic continuation of $G(p)$ given by Eq. (\ref{eq7}) ($\delta$ is an infinitesimally small positive number), at $\mu mR_*^2=0.0298$ and $T mR_*^2=0.0302$ which is just above $T_{\rm c}$ indicated in the inset of Fig.~\ref{fig5}.
One can see that the atomic dispersion has a gap structure near $\omega=0$ even in the absence of the superfluid gap.
This excitation gap in the normal phase originates from strong pairing fluctuations (preformed Cooper pair) and is called pseudogap, which has been extensively discussed for various  strongly correlated quantum systems such as high-$T_{\rm c}$ superconductors \cite{PRL80.149,RMP79.353}, ultracold Fermi gases \cite{Pieri,Perali1,Perali2,Tsuchiya,Tsuchiya2,Watanabe,Perali3,Mueller1,Palestini,Ota,Tajima3,Tajima4,Mueller2,Stewart,Gaebler,Feld,
Magierski,Wlazlowski,Sagi}, color superconductivity \cite{Kitazawa,He2}, and nuclear matter \cite{Schnell,Bozek,Huang}. 
This single-particle excitation property is accessible by photo-emission spectrum measurement in cold atom systems \cite{Stewart,Gaebler,Sagi,Feld}.
In principle, such experiments could also observe many-body effects associated with in-medium Efimov trimers.
However, treating such many-body effects theoretically would require a self-consistent approach including both two-body and three-body correlations in the self-energy.   
\par
\begin{figure}[t]
\begin{center}
\includegraphics[width=8.5cm]{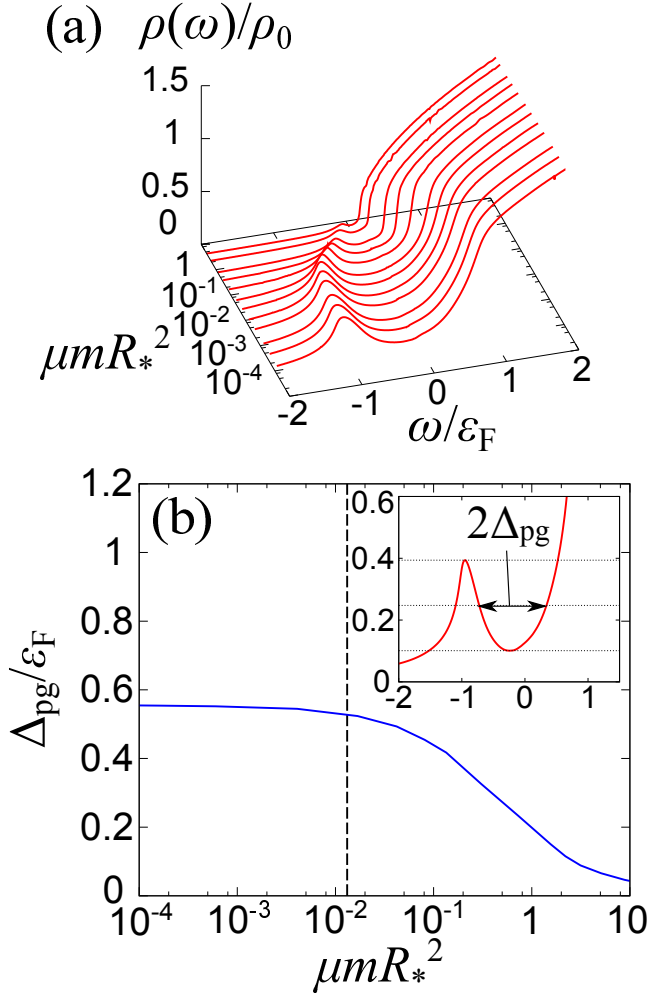}
\end{center}
\caption{(a) Single-particle density of states $\rho(\omega)$ and (b) pseudogap size $\Delta_{\rm pg}$ \cite{Watanabe} at $T=T_{\rm c}$ where $\rho_0$ is that at a Fermi level in a non-interacting Fermi gas.
The vertical dashed line shows $E_3^{\rm M}=0$.
$\Delta_{\rm pg}$ is determined from the half width of the dip in $\rho(\omega)$ around $\omega=0$ as shown in the inset of the panel (b), where the horizontal dotted lines exhibit the local maximum, the half depth, and the local minimum, respectively.
}
\label{fig7}
\end{figure}
The pseudogap can also be seen in the single-particle density of states $\rho(\omega)$, which 
is defined by
\begin{eqnarray}
\label{eq15}
\rho(\omega)=\sum_{\bm{p}}A(\bm{p},\omega).
\end{eqnarray}
It is shown in Fig.~\ref{fig7} (a) at $T=T_{\rm c}$.
This quantity clearly shows the pseudogap effect as a dip structure around $\omega=0$.
The pseudogap disappears away from $\mu=T=0$, that is, in the high-density (large range parameter) limit as found in the case of two-component Fermi gases with negative effective range \cite{Tajima3}.
To characterize this many-body phenomenon, we introduce the pseudogap size $\Delta_{\rm pg}$ defined as the half width of the dip \cite{Watanabe} [see the inset of Fig. \ref{fig7} (b)].
One can find that $\Delta_{\rm pg}$ grows when $\mu mR_*^2$ approaches $0$ and reaches a maximum value $\Delta_{\rm pg}\simeq 0.55\varepsilon_{\rm F}$ at $\mu mR_*^2=0$.
This enhancement of $\Delta_{\rm pg}$ indicates that many-body effects associated with pairing fluctuations are important in the region around $\mu=T=0$ in the phase diagram of Fig. \ref{fig5}.
This confirms the expected competition between formation of Cooper pairs and that of Efimov trimers.
This competition would occur around $E_{3}^{\rm M}=0$, which is shown as the vertical dashed line in Fig.~\ref{fig7}(b).
\par
\section{Summary}
\label{sec4}
In this paper, we have investigated some of the strong-coupling effects occurring in resonantly interacting SU(3) Fermi gases with a finite interaction range, 
namely, the in-medium Efimov trimer and the critical temperature $T_{\rm c}$ of color superfluidity.
\par
The trimer formation is weakened by the medium effects, which consist of thermal agitation and Fermi pressure due to Pauli exclusion.
The trimer is affected by thermal agitation when the thermal de Broglie wavelength
is comparable to the trimer size given by the range of interaction. 
The Pauli-blocking effects are significant when the chemical potential $\mu$ becomes positive.
As in the case of the Gor'kov-Melik-Barkhudarov corrections in weakly coupled superconductors where the pairing interaction forming loosely bound Cooper pairs is screened by electrons from the medium, the medium effects in three-component Fermi gases become stronger when the trimer size or the atomic Fermi sphere becomes larger.
We have shown how these effects would appear in actual experiments varying temperature at fixed number density.
\par 
Finally, we have investigated the phase diagram with respect to the chemical potential $\mu$ and temperature $T$. 
Our calculations indicate the existence of three phases: trimers phase, normal phase, and color superfluidity.
Interestingly, the obtained phase diagram is analogous to the phenomenological QCD phase diagram which consists of hadron, deconfined quark, and color superconducting phases.
We emphasize that the finite interaction range plays an important role to obtain such a QCD-like phase diagram in this atomic system.
\par
Near $\mu=T=0$ in our phase diagram, the system is expected to be dominated by strong two-body and three-body correlations resulting from the competition between trimer formation and color superfluidity.
This idea is supported by our calculation of the in-medium trimer energy and the single-particle spectral function which exhibits strong pairing fluctuations near the color superfluid phase transition.
A self-consistent treatment of two-body and three-body correlations is required to understand this interesting regime, which is left as a future problem.   
Our analysis does not exclude the possibility of trimer superfluidity due to a possible residual attraction between the trimers \cite{Endo3}. 
Such a state would be similar to the $p$-wave superfluidity in a Bose-Fermi mixture \cite{Kinnunen3}.
Although we consider a resonant interaction in this paper, the phase diagram would quantitatively change when tuning the scattering length.
In particular, the color superfluid phase (or molecular BEC) would start at a lower chemical potential in the case of a finite scattering length \cite{Nishida}.
\par
\acknowledgements
H. T. thanks T. Hatsuda, Y. Nishida, and G. Baym for useful discussion.
H. T. was supported by a Grant-in-Aid for JSPS fellows (No.17J03975).
P. N. was supported by the RIKEN Incentive Research Project.
This work was partially supported by RIKEN iTHEMS program.
\par
\appendix
\section{Derivation of the in-medium STM equation}
\label{ApA}
Since we calculate the ground-state trimer energy $E_{3}^{\rm M}$ with the medium effects,
we set $P=(\bm{0},i\zeta_l)$ in the three-body $T$-matrix equation given by Eq. (\ref{eq4}) \cite{Brodsky,Iskin} [where $i\zeta_l=(2l+1)\pi T$ is the fermionic Matsubara frequency].
We obtain
\begin{eqnarray}
\label{eqA1}
T_{3}^{\rm M}(p,p';P)=-2g^2T\sum_{q}G_0(P-p-q)G_0(q)D(P-q)T_3^{\rm M}(q,p';P),
\end{eqnarray}
where we ignore the first term of r. h. s. of Eq. (\ref{eq4}) which is negligible near the pole of $T_3^{\rm M}$.
The summation of the Matsubara frequency $i\Omega_n$ [where $q=(\bm{q},i\Omega_n)$] in Eq. (\ref{eqA1}) can be replaced by the contour integral with respect to an anticlockwise path $C$ \cite{FW} enclosing the pole of the Fermi-Dirac distribution function $f(x)$, namely, $x=i\Omega_n$ as
\begin{eqnarray}
\label{eqA2}
T\sum_{i\Omega_n}=-\oint_C\frac{dx}{2\pi i}f(x),
\end{eqnarray}
since ${\rm Res}f(x=i\Omega_n)=-T$.
We note that $C$ can be deformed to a clockwise path $C'$ which encloses the poles of $G_0$, $D$, and $T_3^{\rm M}$, which give medium effects associated with the momentum distributions of atoms, molecules, and trimers, respectively.
For simplicity, we consider only the pole of $G_0$ to incorporate the effects of the atomic Fermi-Dirac distribution.
This approximation is justified in the high temperature regime where the fugacity $z=e^{\mu/T}$ is small.
In this regime, the atomic Fermi distribution function is approximately given by $f(\xi_{\bm{p}}^{\rm F})\simeq e^{-\frac{p^2}{2mT}}z$, whereas the molecular and trimer distribution functions are $b(\xi_{\bm{Q}}^{\rm B})\simeq e^{-\frac{\bm{Q}^2}{4mT}}z^2$ and
$f(\xi_{\bm{P}}^{\rm T})\simeq e^{-\left(\frac{\bm{P}^2}{6m}+E_3^{\rm M}\right)/T}z^3$, respectively [where $b(\xi)=1/(e^{\xi/T}-1)$ is the Bose-Einstein distribution function and $\xi_{\bm{P}}^{\rm T}=\bm{P}^2/(6m)-3\mu+E_3^{\rm M}$ is the kinetic energy of a trimer].    
Using this approximation, we can analytically perform the energy integration as
\begin{eqnarray}
\label{eqA3}
T_{3}^{\rm M}(p,p';P)&=&2g^2\sum_{\bm{q}}\oint_{C'}\frac{dx}{2\pi i}\frac{f(x)D(P-q)T_3^{\rm M}(q,p';P)}{\left(i\zeta_l-i\omega_n-x-\xi_{\bm{p+\bm{q}}}^{\rm F}\right)\left(x-\xi_{\bm{q}}^{\rm F}\right)}\cr
&=&2g^2\sum_{\bm{q}}
\left[\frac{\left\{1-f(\xi_{\bm{p}+\bm{q}}^{\rm F})\right\}D(P-q_1)T_3^{\rm M}(q_1,p';P)}{i\zeta_l-i\omega_n-\xi_{\bm{q}}^{\rm F}-\xi_{\bm{p+\bm{q}}}^{\rm F}}\right.
\cr
&&-\left.\frac{f(\xi_{\bm{q}}^{\rm F})D(P-q_2)T_3^{\rm M}(q_2,p';P)}{i\zeta_l-i\omega_n-\xi_{\bm{q}}^{\rm F}-\xi_{\bm{p+\bm{q}}}^{\rm F}}\right],
\end{eqnarray}
where $q_1=(\bm{q},i\zeta_l-i\omega_n-\xi_{\bm{p}+\bm{q}}^{\rm F})$ and $q_2=(\bm{q},\xi_{\bm{q}}^{\rm F})$.
To obtain the in-medium STM equation, we perform the analytic continuations  $i\omega_n\rightarrow \xi_{\bm{p}}^{\rm F}$ and $i\zeta_l\rightarrow E_{3}^{\rm M}-3\mu$ in Eq. (\ref{eqA3}).
In this way, we obtain
\begin{eqnarray}
\label{eqA4}
T_{3}^{\rm M}(p,p';P)&=&2g^2\sum_{\bm{q}}
\left[\frac{\left\{1-f(\xi_{\bm{p}+\bm{q}}^{\rm F})\right\}D(\bm{q},\xi_{\bm{p}+\bm{q}}^{\rm F}+\xi_{\bm{p}}^{\rm F})}{E_3^{\rm M}-3\mu-\xi_{\bm{p}}^{\rm F}-\xi_{\bm{q}}^{\rm F}-\xi_{\bm{q}+\bm{p}}^{\rm F}}\right.
\cr
&&\times T_3^{\rm M}\left(\left\{\bm{q},E_3^{\rm M}-3\mu-\xi_{\bm{p}}^{\rm F}-\xi_{\bm{p}+\bm{q}}^{\rm F}\right\},p';P\right)
\cr
&&-\left.\frac{f(\xi_{\bm{q}}^{\rm F})D(\bm{q},E_3^{\rm M}-3\mu-\xi_{\bm{q}}^{\rm F})T_3^{\rm M}(\{\bm{q},\xi_{\bm{q}}^{\rm F}\},p';P)}{E_3^{\rm M}-3\mu-\xi_{\bm{p}}^{\rm F}-\xi_{\bm{q}}^{\rm F}-\xi_{\bm{q}+\bm{p}}^{\rm F}}\right].
\end{eqnarray}
Furthermore, the first integrand in Eq. (\ref{eqA4}) gives dominant contributions near its pole, corresponding to $E_3^{\rm M}-3\mu-\xi_{\bm{p}}^{\rm F}-\xi_{\bm{q}}^{\rm F}-\xi_{\bm{q}+\bm{p}}^{\rm F}=0$.
We can then approximate the arguments of $D$ and $T_3^{\rm M}$ as
\begin{eqnarray}
\label{eqAadd1}
D(\bm{q},\xi_{\bm{p}+\bm{q}}^{\rm F}+\xi_{\bm{p}}^{\rm F})\simeq D(\bm{q},E_3^{\rm M}-3\mu-\xi_{\bm{q}}^{\rm F}),
\end{eqnarray}
and
\begin{eqnarray}
\label{eqAadd2}
T_3^{\rm M}\left(\left\{\bm{q},E_3^{\rm M}-3\mu-\xi_{\bm{p}}^{\rm F}-\xi_{\bm{p}+\bm{q}}^{\rm F}\right\},p';P\right)\simeq T_3^{\rm M}\left(\left\{\bm{q},\xi_{\bm{q}}^{\rm F}\right\},p';P\right).
\end{eqnarray}
By substituting Eqs. (\ref{eqAadd1}) and (\ref{eqAadd2}) into Eq. (\ref{eqA4}),
we obtain
\begin{eqnarray}
\label{eqA4d}
T_{3}^{\rm M}(p,p';P)&=&2g^2\sum_{\bm{q}}
\frac{1-f(\xi_{\bm{q}}^{\rm F})-f(\xi_{\bm{p}+\bm{q}}^{\rm F})}{E_3^{\rm M}-3\mu-\xi_{\bm{p}}^{\rm F}-\xi_{\bm{q}}^{\rm F}-\xi_{\bm{q}+\bm{p}}^{\rm F}}
\cr
&&\times D(\bm{q},E_3^{\rm M}-3\mu-\xi_{\bm{q}}^{\rm F}) T_3^{\rm M}\left(\left\{\bm{q},\xi_{\bm{q}}^{\rm F}\right\},p';P\right),
\end{eqnarray}
where
\begin{eqnarray}
\label{eqA5}
D(\bm{q},E_3^{\rm M}-3\mu-\xi_{\bm{q}}^{\rm F})=\frac{4\pi}{mg^2}
\left[-R_*\kappa(\bm{q})^2+4\pi\sum_{\bm{k}}\left\{\frac{F(\bm{k},\bm{q})}{\bm{k}^2+\kappa(\bm{q})^2}-\frac{1}{\bm{k}^2}\right\}\right]^{-1},
\end{eqnarray}
is obtained from the analytic continuation of Eq. (\ref{eq9}).
We note that $\kappa(\bm{q})^2=\frac{3}{4}\bm{q}^2-mE_3^{\rm M}$
and $F(\bm{k},\bm{q})$ is the statistical factor defined by Eq. (\ref{eq6}).
We introduce
\begin{eqnarray}
\label{eqA6}
L(\bm{q})=\frac{mg^2}{4\pi}D(\bm{q},E_3^{\rm M}-3\mu-\xi_{\bm{q}}^{\rm F})T_3^{\rm M}(\{\bm{q},\xi_{\bm{q}}^{\rm F}\},p';P).
\end{eqnarray}
We note that although $L(\bm{q})$ implicitly depends on $p'$ and $P$,
they do not change the equation of $E_3^{\rm M}$.
By using $L(\bm{q})$, Eq. (\ref{eqA4d}) can be rewritten by
\begin{eqnarray}
\label{eqA7}
&&\left[-R_*\kappa(\bm{p})^2+4\pi\sum_{\bm{k}}\left\{\frac{F(\bm{k},\bm{p})}{\bm{k}^2+\kappa(\bm{p})^2}-\frac{1}{\bm{k}^2}\right\}\right]L(\bm{p}) \cr
&&=8\pi\sum_{\bm{q}}\frac{1-f(\xi_{\bm{p}+\bm{q}}^{\rm F})-f(\xi_{\bm{q}}^{\rm F})}{mE_3^{\rm M}-\bm{p}^2-\bm{q}^2-\bm{p}\cdot\bm{q}}L(\bm{q}).
\end{eqnarray}
Finally, we obtain the in-medium STM equation, that is, Eq. (\ref{eq5}) by making the substitutions $\bm{q}\rightarrow \bm{q}+\bm{p}/2$ and $\bm{p}\rightarrow -\bm{p}$ in Eq. (\ref{eqA7}).
We note that by taking the vacuum limit $\mu\rightarrow -\infty$ where $f(\xi_{\bm{p}}^{\rm F})\rightarrow 0$, Eq. (\ref{eq5}) reproduces the ordinary STM equation of a three-body problem at $1/a=0$ \cite{Nishida} given by
\begin{eqnarray}
\label{eqA8}
\left[-R_*\kappa(\bm{p})^2+4\pi\sum_{\bm{k}}\left\{\frac{1}{\bm{k}^2+\kappa(\bm{p})^2}-\frac{1}{\bm{k}^2}\right\}\right]L\left(\bm{p}\right)=-8\pi\sum_{\bm{q}}\frac{L(\bm{q+\bm{p}/2})}{\bm{q}^2+\kappa(\bm{p})^2},
\end{eqnarray}
which gives the ground-state trimer energy $E_3^{\rm V}=-0.01385/(mR_*^2)$ in vacuum.

\end{document}